\documentclass[12pt]{iopart}

\usepackage{graphicx}
\begin{document}

\title{Quadrupole and octupole states in $^{152}$Sm using the proton-neutron interacting boson model}

\author{Bao-Yue Hu}
\address{Department of Physics, Liaoning Normal University,
Dalian 116029, China}

\author{Yu Zhang*}
\address{Department of Physics, Liaoning Normal University,
Dalian 116029, China}

\author{Gui-Xiu Na}
\address{Department of Physics, Liaoning Normal University,
Dalian 116029, China}

\author{Sheng-Nan Wang}
\address{Department of Physics, Liaoning Normal University,
Dalian 116029, China}

\author{Wei Teng}
\address{Department of Physics, Liaoning Normal University,
Dalian 116029, China}

\ead{dlzhangyu\_physics@163.com}

\begin{abstract}
A scheme of solving the proton-neutron interacting boson model
(IBM-2) in terms of the SU(3) basis is introduced, by
which the IBM-2 coupled with an octupole boson is
applied to describe the low-energy structure of the critical
point nucleus, $^{152}$Sm. The results indicate that the spectral properties of
both the positive-parity bands and negative-parity bands in this nucleus can be
well captured by the IBM-2 calculations through a simple Hamiltonian, thus providing an example
of the IBM-2 in a unified description of quadrupole and octupole states in a transitional system.
In addition, a statistical analysis of the low-spin states in the model is also provided.
\end{abstract}

\maketitle

\section{Introduction}

The interacting boson model (IBM)~\cite{Iachellobook} has achieved
great success in exploring low-energy structural properties of heavy and intermediately-heavy
nuclei. The model has two versions,
the IBM-1 of treating proton and neutron as the same type of
identical particle and the IBM-2 that distinguishes proton from
neutron, also called the proton-neutron interacting boson model~\cite{Iachellobook}.
The IBM-2 has a robust shell-model
foundation~\cite{IachelloTalmi} with $s,~d$ bosons being explained from the $l^P=0^+,~2^+$ pairs of valence nucleons~\cite{Otsuka1981}.
Specifically, the model can be introduced as a drastic truncation of large scale shell model calculations with a boson mapping~\cite{Otsuka1978,Otsuka1978II}.
The widely used consistent-$Q$ Hamiltonian form in the IBM-2 is directly motivated from
the shell model Hamiltonian~\cite{IachelloTalmi}. Recent years, a systematic
application of the IBM-2 has been conducted~\cite{Nomura2008,Nomura2011,Nomura2011II,Nomura2011III,Nomura2011IV,Nomura2017,Nomura2019}
with the model parameters being self-consistently determined from
the microscopic mean-field calculations, which further emphasizes
the microscopic aspect of the IBM-2. In addition,
the IBM-2 was also employed to calculate nuclear matrix elements of
neutrinoless double-$\beta$ decay~\cite{Barea2012,Barea2013} in the mass
region that is out of reach of shell model calculations. All these
demonstrate that the IBM-2 can provide a comprehensive and meanwhile quantitatively good
framework for exploring various aspects of nuclear systems.

The transitional nucleus, $^{152}$Sm, due to its rich structural
features~\cite{Martin2013} was often chosen to test theoretical
predictions of different
models~\cite{Iachello1998,Iachello2001,Casten2001,Jolie1999,Li2009,Gupta2017,Heyde2011}.
Particularly, this nucleus can be regarded as an experimental candidate of the critical point
symmetry~\cite{Iachello2001,Casten2001} for the spherical-deformed shape phase
transition~\cite{CJC2010}. On the theoretical side, the IBM-2 with rich phase structures offers a convenient
frame to study shape phase transitions~\cite{Caprio2004,Kotila2012}. Nevertheless, the previous IBM-2
analysis of this nucleus were mostly confined to positive-parity
states~\cite{Iachellobook,Nomura2011IV,Nomura2019}. In
experiments~\cite{Martin2013}, the rotational bands formed by the
negative-parity states in $^{152}$Sm have been clearly observed and suggested
to be associated with octupole vibration~\cite{Konljin1982}.
Undoubtedly, if quadrupole and
octupole states are simultaneously involved in theoretical calculation,
it would place a higher requirement and more rigorous test on the model. A way of generating octupole states in the IBM is
to extend the model by including $f$ boson with $l^P=3^-$. There
exist few works with the octupole degree of freedom considered
in the IBM-2 frame. In \cite{Yoshinaga1993}, the $sdfg$-IBM-2 was
applied to describe the Ra isotopes with the Hamiltonian being
solved through projecting it to the IBM-1. In
\cite{Smirnova2000,Pietralla2003}, the $sdf$-IBM-2 was discussed with the
emphasis placed on the dynamical symmetry situations. Very recently, a more general $sdf$-IBM-2 Hamiltonian has been employed to give a systematical investigation of octupole correlations in the collective states of neutron-rich nuclei with $N_\mathrm{n}\approx56$~\cite{Nomura2022}. In the model, all the interactional strengthes can be fixed by a self-consistent mean-field calculation using the microscopic energy density functional~\cite{Nomura2022}. In addition, a systematic study of the spectroscopic properties of the Rn isotopes was
recently carried in the $spdf$-IBM-2~\cite{Vallejos2021}, in which
negative-parity and positive-parity configurations are treated on equal footing.
Comparatively, most studies of octupole states were
performed in the IBM-1
frame~\cite{Scholten1978,Barfield1986,Barfield1988,Engel1987,Otsuka1988,Zamfir2001,Nomura2013,Nomura2014,Nomura2015},
where the model space for a given nucleus may be much smaller than the
corresponding IBM-2 version.

There are two folds in this work. First we would like introduce a
diagonalization scheme based on the weak-coupling SU(3) basis to
solve the Hamiltonian in the IBM-2 and its extension. In contrast, the traditionally used algorithm is
called NPBOS, by which the IBM-2 Hamiltonian is solved in terms of the U(5) basis~\cite{NPBOS}. Second, with the numerical
scheme, we will present a unified IBM-2 description of the positive-parity and negative-parity states in $^{152}$Sm to examine the two-fluid model explanation of a transitional system.

\section{The Model}

There are two kinds of boson in the IBM, the monopole $s$ boson with
$l^P=0^+$ and the quadrupole $d$ boson with $l^P=2^+$. In the IBM-2, the degree
of freedom of proton will be distinguished from that of neutron,
which means that proton boson and neutron boson are two different
types of boson. For proton (neutron) boson, the bilinear products of
creation and annihilation operators,
\begin{eqnarray}
G_{\alpha\beta}^{(\rho)}=b^\dag_{\rho,\alpha}b_{\rho,\beta},~(\alpha,\beta=1,\cdots,6;~\rho=\pi,\nu)\, ,
\end{eqnarray}
include totally 36 independent elements, which generate the dynamical
symmetry $\mathrm{U}(6)$. As a result, the proton-neutron
coupling system (IBM-2) has the direct product dynamical
symmetry $\mathrm{U}_\pi(6)\otimes \mathrm{U}_\nu(6)$. There are
different ways in which the $\mathrm{U}_\pi(6)\otimes
\mathrm{U}_\nu(6)$ group can be reduced to
$\mathrm{SO}_{\pi+\nu}(3)\supset\mathrm{SO}_{\pi+\nu}(2)$~\cite{Iachellobook}.
Each of them represents a dynamical symmetry limit and offers a set of complete basis to solve the IBM-2
Hamiltonian.

\subsection{The SU(3) Basis}

In this work, we focus on the basis associated with the SU(3) symmetry limit which is
characterized by the group chain
\begin{eqnarray}&&\mathrm{U}_\pi(6)\otimes \mathrm{U}_\nu(6)\supset \mathrm{SU}_\pi(3)\otimes
\mathrm{SU}_\nu(3)\\ \nonumber
&&~~~~\supset\mathrm{SO}_\pi(3)\otimes
\mathrm{SO}_\nu(3)\supset\mathrm{SO}_{\pi+\nu}(3)\supset
\mathrm{SO}_{\pi+\nu}(2)\, .\end{eqnarray} The corresponding SU(3) basis states in the coupled
system can be constructed as
\begin{eqnarray}\label{basis}
&|\alpha LM\rangle\equiv|N_\pi(\lambda_\pi,\mu_\pi)\chi_\pi
L_\pi;N_\nu(\lambda_\nu,\mu_\nu)\chi_\nu L_\nu;LM\rangle\\ \nonumber
&=\sum_{M_\pi( M_\nu)}\langle
L_\pi M_\pi L_\nu M_\nu|LM\rangle|N_\pi(\lambda_\pi,\mu_\pi)\chi_\pi
L_\pi
M_\pi\rangle|N_\nu(\lambda_\nu,\mu_\nu)\chi_\nu L_\nu
M_\nu\rangle\,
\end{eqnarray}
with $\langle~\mid\rangle$ denoting the SO(3) CG-coefficients. In Eq.~(\ref{basis}),
$N_\rho,~(\lambda_\rho,~\mu_\rho),~L_\rho$ and $M_\rho$ with $\rho=\pi,\nu$ represent
the quantum numbers that are used to signify the irreducible representations (IRREPs) of
$\mathrm{U}_\rho(6),~\mathrm{SU}_\rho(3),~\mathrm{SO}_\rho(3)$ and
$\mathrm{SO}_\rho(2)$, respectively, while $\chi_\rho$ stands for the additional
quantum number in the reduction
$\mathrm{SU}_\rho(3)\supset\mathrm{SO}_\rho(3)$. Once the boson
number $N_\rho$ is known, the SU(3) IRREPs,
$(\lambda_\rho,\mu_\rho)$, are given by
\begin{eqnarray}\nonumber
(\lambda,\mu)&=&(2N,0)\oplus(2N-4,2)\oplus(2N-8,4)\oplus\cdots\\
\nonumber
&~&\oplus(2N-6,0)\oplus(2N-10,2)\oplus\cdots\\
\nonumber
&~&\oplus(2N-12,0)\oplus(2N-16,2)\oplus\cdots\\
&&~~~~~~~~~~~~~~~~~~\cdots\, .
\end{eqnarray}
For each $(\lambda,\mu)$, the quantum number $L$ is
determined by
\begin{eqnarray}
&&K=\mathrm{min}({\lambda,\mu}),\mathrm{min}({\lambda,\mu})-2,\cdots,0,\\
\nonumber &&K>0,~~L=K,K+1,\cdots,K+\mathrm{max}({\lambda,\mu}),\\
\nonumber &&K=0,~~L=0,~2,~4,\cdots,~\mathrm{max}({\lambda,\mu})\, ,
\end{eqnarray}
where $K$ is the projection of $L$ on the $z$-axis of the intrinsic
frame. Note that the basis states with $K$ as the additional quantum number
are not orthogonal to each other. Instead, we adopt here the orthogonal
Draayer-Akiyama SU(3) basis~\cite{DraayerAkiyama1973} as given in Eq.~(\ref{basis}).

With $s~d$ bosons, the IBM-2 can only describe
quadrupole states of positive parity. If octupole states of
negative parity are taken into account, a
simple manner is to add an $f$ boson with $l^P=3^-$ to the usual
$s$-$d$ boson space~\cite{Scholten1978,Barfield1986,Barfield1988}.
Accordingly, the SU(3) basis for octupole states can be built from
the basis (\ref{basis}) and given by
\begin{eqnarray}\label{basisII}\nonumber
|(\alpha L) JM_J\rangle&\equiv&\Big[|\alpha
L\rangle\otimes|n_fL_f\rangle\Big]_{M_J}^J\\
&=&\sum_{M_f(M)}\langle L M L_f M_f|JM_J\rangle|\alpha
LM\rangle|n_fL_fM_f\rangle\,
\end{eqnarray}
with $n_f=1$ and $L_f=3$.
Note that $f$ boson in the IBM-2 can be either a
proton boson or a neutron boson. It means that the SU(3) basis for octupole states
in (\ref{basisII}) should include two sets, $(N_\pi-1,~N_\nu,~n_{f_\pi}=1)$ and $(N_\pi,~N_\nu-1,~n_{f_\nu}=1)$,
due to the boson number conservation.
Clearly, the basis vectors in the two sets are
orthogonal to each other. If only quadrupole states are concerned,
one can set $n_f=0$, by which the SU(3) basis
(\ref{basisII}) will be directly reduced back to that given in (\ref{basis}). In other words, the SU(3)
basis defined in (\ref{basisII}) have covered both quadrupole states
with $n_f=0$ and octupole states with $n_f=1$.
Microscopically, the
octupole characters in nuclear system are connected to the octupole correlation of
a pair of single-particle orbits with opposite parities. The orbit pairs satisfying the related conditions are
$(h_{11/2},d_{5/2})$ in the 50-82 shell and $(i_{13/2},f_{7/2})$ in
the 82-126 shell~\cite{Barfield}. The 2qp configurations generated by
these orbit pairs are expected to be important for octupole band at
least in the well deformed cases~\cite{Barfield}. In this work, we
do not emphasize the microscopic origin of octupole boson but
follow the IBM-1+$f$ model to construct the corresponding IBM-2
counterpart.

\subsection{Matrix Elements}

The creation and annihilation operators of $s,~d$ boson can be expressed as the
$\mathrm{SU(3)}$ tensors, $A^{(2,0)}$ and
$B^{(0,2)}$~\cite{Rosensteel1990}, by which one can calculate the matrix
elements of any boson operators in the SU(3) basis defined in
(\ref{basis}) and (\ref{basisII}). In what follows, we take
three types of neutron-proton (NP) interaction as the examples to illustrate how to
calculate the Hamiltonian matrix. They are
\begin{eqnarray}\label{QQI}
&&\hat{Q}_d^\pi\cdot\hat{Q}_d^\nu=(d_\pi^\dag\times\
\tilde{d_\pi})^{(2)}\cdot(d_\nu^\dag\times \tilde{d_\nu})^{(2)},\\
&&\label{QQII} \hat{Q}_d^\pi\cdot\hat{Q}_f^\nu=(d_\pi^\dag\times
\tilde{d_\pi})^{(2)}\cdot(f_\nu^\dag\times \tilde{f_\nu})^{(2)},\\
\label{QQIII}
&&\hat{O}_{df}^\pi\cdot\hat{O}_{df}^\nu=(d_\pi^\dag\times
\tilde{f_\pi})^{(3)}\cdot(f_\nu^\dag\times
\tilde{d_\nu})^{(3)}+\mathrm{h.c.}\, ,
\end{eqnarray} where the spherical tensor form $\tilde{b}_m^l=(-1)^{l-m}b_{-m}$ is used and $\mathrm{h.c.}$ represents the
Hermitian conjugate. Based on the Wigner-Eckart theorem, one can derive that
\begin{eqnarray}\nonumber\label{A}
&&\langle\alpha_bLM\mid\hat{Q}_d^\pi\cdot\hat{Q}_d^\nu\mid\alpha_aLM\rangle\\
\nonumber
&&=5(-1)^{L_{\pi b}+L_{\nu
a}+L}\sum_{\lambda_{1,2}\mu_{1,2}\chi_{1,2}L_{1,2}}\left\{\begin{array}{ccc}L_{\pi
b}~L_{\nu b}~L
\\ \nonumber L_{\nu a}~L_{\pi a}~2\end{array}
\right\} \left\{\begin{array}{ccc}2~~~~2~~~~2
\\ \nonumber L_{\pi a}~L_{\pi b}~L_1\end{array}
\right\}\\
\nonumber
&&\times\left\{\begin{array}{ccc}2~~~~2~~~~2
\\ \nonumber L_{\nu a}~L_{\nu b}~L_2\end{array} \right\}\sqrt{2L_{\pi b}+1}\sqrt{2L_1+1}\sqrt{2L_{\nu
b}+1}\sqrt{2L_2+1}\\ \nonumber
&&\times\langle(\lambda_1,\mu_1)\chi_1L_1;(20)12\parallel(\lambda_{\pi
b},\mu_{\pi b})\chi_{\pi b}L_{\pi b}\rangle\\ \nonumber
&&\times\langle(\lambda_{\pi a},\mu_{\pi a})\chi_{\pi a}L_{\pi
a};(02)12\parallel(\lambda_{1},\mu_{1})\chi_{1}L_{1}\rangle\\
\nonumber
&&\times\langle(\lambda_2,\mu_2)\chi_2L_2;(20)12\parallel(\lambda_{\nu
b},\mu_{\nu b})\chi_{\nu b}L_{\nu b}\rangle\\
\nonumber
&&\times\langle(\lambda_{\nu a},\mu_{\nu a})\chi_{\nu
a}L_{\nu
a};(02)12\parallel(\lambda_{2},\mu_{2})\chi_{2}L_{2}\rangle\\
\nonumber
&&\times\langle N_\pi(\lambda_{\pi b},\nu_{\pi
b})\parallel\mid A^{(20)}\parallel\mid
N_\pi-1(\lambda_1,\mu_1)\rangle\\ \nonumber
&&\times\langle
N_\pi-1(\lambda_{1},\mu_{1})\parallel\mid B^{(02)}\parallel\mid
N_\pi(\lambda_{\pi a},\mu_{\pi a})\rangle\\ \nonumber
&&\times\langle
N_\nu(\lambda_{\nu b},\mu_{\nu b})\parallel\mid A^{(20)}\parallel\mid
N_\nu-1(\lambda_2,\mu_2)\rangle\\
&&\times\langle
N_\nu-1(\lambda_{2},\nu_{2})\parallel\mid B^{(02)}\parallel\mid
N_\nu(\lambda_{\nu a},\mu_{\nu a})\rangle\, .
\end{eqnarray}
It is shown that the matrix element can be decomposed into the
product of the reduced matrix elements of single-boson operator and
the relevant coefficients. In (\ref{A}), the $\mathrm{SU(3)}\supset
\mathrm{SO(3)}$ isoscalar factors
$\langle~\parallel\rangle$ can be calculated using the
Draayer-Akiyama algorithm~\cite{DraayerAkiyama1973}, the three bar
matrix elements, $\langle~\mid\parallel A^{(20)}\parallel\mid~\rangle$
and $\langle~\mid\parallel B^{(02)}\parallel\mid~\rangle$, can be
found in \cite{Rosensteel1990}, while  $\{~~\}$ are the SO(3) 6-$j$
coefficients. Clearly, the quadrupole-quadrupole interaction defined in
(\ref{QQI}) may have contributions to both quadrupole and octupole states. It is
easy to deduce that
\begin{eqnarray}\label{quadrupoleoctupole}
&&\langle(\alpha_bL_b)JM_J|\hat{Q}_d^\pi\cdot\hat{Q}_d^\nu|\langle(\alpha_aL_a)JM_J\rangle\\
\nonumber
&&~~~~~~~~~~~~=\langle\alpha_bL_bM_b|\hat{Q}_d^\pi\cdot\hat{Q}_d^\nu|\langle\alpha_aL_aM_a\rangle\delta_{L_a,L_b}\delta_{M_a,M_b}\,
\end{eqnarray} with
$|L_a-3|\leq J\leq L_a+3$, which means that this term contributes
equally for quadrupole and octupole states except that one should adopt $N_\rho-1$ instead of $N_\rho$ in the calculations
for octuple states due to the boson number conservation for a given nucleus.

The interaction terms involving $f$-boson in the present scheme will only contribute
to octupole states. Similarly, one can derive
\begin{eqnarray}\nonumber
&&\langle(\alpha_bL_b)JM_J\mid\hat{Q}_d^\pi\cdot\hat{Q}_f^\nu \mid(\alpha_aL_a)JM_J\rangle\\
\nonumber &&=-\delta_{1,~n_{f_\nu}}5(-1)^{L_{\pi a}+L_{\nu
a}+J}\sum_{\lambda_{1}\mu_{1}\chi_{1}L_{1}}\left\{\begin{array}{ccc}L_{
b}~~~~3~~~~J
\\ \nonumber ~3~~~~L_{a}~~~~2\end{array}
\right\} \left\{\begin{array}{ccc}L_{\pi b}~~L_{\pi a}~~2
\\ \nonumber ~L_{a}~~~L_{b}~~L_{\nu a}\end{array}
\right\}\\ \nonumber
&&\times\left\{\begin{array}{ccc}2~~~~2~~~~2
\\ \nonumber L_{\pi a}~L_{\pi b}~L_1\end{array} \right\}\sqrt{2L_{a}+1}\sqrt{2L_b+1}\sqrt{2L_{\pi b}+1}\sqrt{2L_1+1}\\
\nonumber
&&\times\langle(\lambda_1,\mu_1)\chi_1L_1;(20)12\parallel(\lambda_{\pi
b},\mu_{\pi b})\chi_{\pi b}L_{\pi b}\rangle\\ \nonumber
&&\times\langle(\lambda_{\pi a},\mu_{\pi a})\chi_{\pi a}L_{\pi
a};(02)12\parallel(\lambda_{1},\mu_{1})\chi_{1}L_{1}\rangle\\
&&\times\langle N_\pi(\lambda_{\pi b},\mu_{\pi b})\parallel\mid
A^{(20)}\parallel\mid N_\pi-1(\lambda_1,\mu_1)\rangle\\ \nonumber
&&\times\langle N_\pi-1(\lambda_{1},\mu_{1})\parallel\mid
B^{(02)}\parallel\mid N_\pi(\lambda_{\pi a},\mu_{\pi a})\rangle\, .
\end{eqnarray}
It is shown that the matrix element can be also decomposed into the
product of the reduced matrix elements of single $d$-boson
operator and the related coefficients since the matrix element of
single $f$-boson operator have been directly worked out for $n_f=1$.
In the same way, one can obtain
\begin{eqnarray}\nonumber
&&\langle(\alpha_bL_b)JM_J\mid\hat{O}_{df}^\pi\cdot\hat{O}_{df}^\nu \mid(\alpha_aL_a)JM_J\rangle\\
\nonumber &=&\delta_{1,n_{f_{\nu b}}}\delta_{1,n_{f_{\pi
a}}}7\sum_{L_\lambda}(-1)^{L_{a}+J+L_\lambda}\left\{\begin{array}{ccc}L_{\pi
b}~~~~L_{\pi a}~~~~2
\\ \nonumber L_{\nu
b}~~~~L_{\nu a}~~~~2\\ \nonumber
~~L_{b}~~~~~~L_{a}~~~~L_\lambda\end{array}
\right\}\\
\nonumber &\times& \left\{\begin{array}{ccc}L_b~~~3~~~J
\\ \nonumber ~3~~~L_a~~~L_{\lambda}\end{array}
\right\}\left\{\begin{array}{ccc}~2~~~~2~~~~L_\lambda
\\ \nonumber 3~~~~3~~~~3\end{array} \right\}(2L_\lambda+1)\\ \nonumber
&\times&\sqrt{2L_b+1}\sqrt{2L_a+1}\sqrt{2L_{\pi b}+1}\sqrt{2L_{\nu b}+1}\\
\nonumber &\times&\langle(\lambda_{\nu a},\mu_{\nu a})\chi_{\nu
a}L_{\nu a};(02)12\parallel(\lambda_{\nu b},\mu_{\nu b})\chi_{\nu
b}L_{\nu b}\rangle\\ \nonumber &\times&\langle(\lambda_{\pi
a},\mu_{\pi a})\chi_{\pi a}L_{\pi a};(20)12\parallel(\lambda_{\pi
b},\mu_{\pi b})\chi_{\pi b}L_{\pi b}\rangle\\
&\times&\langle N_\pi(\lambda_{\pi b},\mu_{\pi b})\parallel\mid
A^{(20)}\parallel\mid N_\pi-1(\lambda_{\pi a},\mu_{\pi a})\rangle\\
\nonumber &\times&\langle N_\nu-1(\lambda_{\nu b},\mu_{\nu
b})\parallel\mid B^{(02)}\parallel\mid N_\nu(\lambda_{\nu a},\mu_{\nu
a})\rangle\\ \nonumber &+&\delta_{1,n_{f_{\pi
b}}}\delta_{1,n_{f_{\nu
a}}}7\sum_{L_\lambda}(-1)^{L_{a}+J}\left\{\begin{array}{ccc}L_{\pi
b}~~~~L_{\pi a}~~~~2
\\ \nonumber L_{\nu
b}~~~~L_{\nu a}~~~~2\\ \nonumber
~~L_{b}~~~~~~L_{a}~~~~L_\lambda\end{array}
\right\}\\
\nonumber &\times& \left\{\begin{array}{ccc}L_{b}~~~3~~~J
\\ \nonumber ~3~~~L_{a}~~L_{\lambda}\end{array}
\right\}\left\{\begin{array}{ccc}2~~~2~~~L_\lambda
\\ \nonumber 3~~~3~~~3\end{array} \right\}(2L_\lambda+1)\\ \nonumber
&\times&\sqrt{2L_b+1}\sqrt{2L_a+1}\sqrt{2L_{\pi b}+1}\sqrt{2L_{\nu b}+1}\\
\nonumber &\times&\langle(\lambda_{\nu a},\mu_{\nu a})\chi_{\nu
a}L_{\nu a};(20)12\parallel(\lambda_{\nu b},\mu_{\nu b})\chi_{\nu
b}L_{\nu b}\rangle\\ \nonumber &\times&\langle(\lambda_{\pi
a},\mu_{\pi a})\chi_{\pi a}L_{\pi a};(02)12\parallel(\lambda_{\pi
b},\mu_{\pi b})\chi_{\pi b}L_{\pi b}\rangle\\
\nonumber&\times&\langle N_\nu(\lambda_{\nu b},\mu_{\nu
b})\parallel\mid
A^{(20)}\parallel\mid N_\nu-1(\lambda_{\nu a},\mu_{\nu a})\rangle\\
\nonumber &\times&\langle N_\pi-1(\lambda_{\pi b},\mu_{\pi
b})\parallel\mid B^{(02)}\parallel\mid N_\pi(\lambda_{\pi a},\mu_{\pi
a})\rangle\, .
\end{eqnarray}
Although this term seems more complicate than the others, the matrix element
can be also attributed to the calculations for the reduced elements of single-boson operator and the relevant
coefficients including the SO(3) 6-$j$ and 9-$j$ ones.
The above examples demonstrate the routine of constructing the
Hamiltonian matrix in the SU(3) basis. With this scheme, one can solve the IBM-2 or IBM-2+$f$ Hamiltonian involving any interactional terms.

\subsection{The $\mathrm{O}_{\pi+\nu}(6)$ Symmetry}

Spectrum in a dynamical symmetry limit can be obtained in an analytical way,
which offers a chance to examine the SU(3) scheme described above.
In the following, the $\mathrm{O}_{\pi+\nu}(6)$ symmetry limit
characterized by
\begin{eqnarray}\label{O6}\nonumber
&&\mathrm{U}_\pi(6)\otimes \mathrm{U}_\nu(6) \supset
\mathrm{U}_{\pi+\nu}(6)\supset
\mathrm{O}_{\pi+\nu}(6)\\
&&~~~~~~~~~~\supset\mathrm{O}_{\pi+\nu}(5)\supset\mathrm{SO}_{\pi+\nu}(3)\supset
\mathrm{SO}_{\pi+\nu}(2)\, \end{eqnarray} is taken as an example to test this numerical scheme of solving the IBM-2 Hamiltonian.
The Hamiltonian for the
$\mathrm{O}_{\pi+\nu}(6)$ dynamical symmetry can be constructed as a linear combination
of the Casimir operators $\hat{C}_k[\mathrm{G}]$~\cite{Iachellobook} for the groups G involved
in the group chain (\ref{O6}). For simplicity, we chose only the quadrupole-quadrupole interaction and the Hamiltonian can be
written as
\begin{eqnarray}\label{O6H}\nonumber
\hat{H}_{\mathrm{O(6)}}&=&B~\hat{Q}_{\pi+\nu}\cdot\hat{Q}_{\pi+\nu}\\
&=&B~\Big(\hat{C}_2[\mathrm{O}_{\pi+\nu}(6)]-\hat{C}_2[\mathrm{O}_{\pi+\nu}(5)]\Big)
\, \end{eqnarray}
with $B<0$ denoting the interactional strength. In (\ref{O6H}), the $\mathrm{O}_{\pi+\nu}(6)$ quadrupole
operator is defined by \begin{eqnarray}
\hat{Q}_{\pi+\nu}=(s_\pi^\dag\times\tilde{d}_\pi+ d_\pi^\dag\times
\tilde{s}_\pi)^{(2)}+(s_\nu^\dag\times\tilde{d}_\nu+ d_\nu^\dag\times
\tilde{s}_\nu)^{(2)}\, .\end{eqnarray} Then, the eigenvectors for
quadrupole states can be expressed as~\cite{Iachellobook}
\begin{eqnarray}\label{O6vector}
|\Psi\rangle=\mid[N_\pi],[N_\nu],(N_1,N_2),(\sigma_1,\sigma_2),(\tau_1,\tau_2),\tilde{n}_{1\Delta},~\tilde{n}_{2\Delta};
LM\rangle\, ,
\end{eqnarray}
in which $(N_1,N_2),(\sigma_1,\sigma_2),(\tau_1,\tau_2)$ are the
quantum numbers for $\mathrm{U}_{\pi+\nu}(6)$,
$\mathrm{O}_{\pi+\nu}(6)$ and $\mathrm{O}_{\pi+\nu}(5)$,
respectively, with $\tilde{n}_{1\Delta},~\tilde{n}_{2\Delta}$ denoting the additional quantum numbers in the reduction
$\mathrm{O}_{\pi+\nu}(5)\supset\mathrm{SO}_{\pi+\nu                                                                                                                                                                              }(3)$. Accordingly, the eigenvalues for
the $\mathrm{O}_{\pi+\nu}(6)$ Hamiltonian can be analytical
given as
\begin{eqnarray}\label{o6II}
E=B~[\sigma_1(\sigma_1+4)+\sigma_2(\sigma_2+2)-\tau_1(\tau_1+3)-\tau_2(\tau_2+1)]\, .
\end{eqnarray} For yrast states, it is given by
$\sigma_1=N_1=N_\pi+N_\nu$, $\tau_2=\sigma_2=N_2=0$ and $L=2\tau_1$
with $\tau_1=0,~1,~2,...,\sigma_1$~\cite{Iachellobook}.

On the other side, the $\mathrm{O}_{\pi+\nu}(6)$ Hamiltonian
(\ref{O6H}) can be solved in a numerical way by using
the SU(3) scheme proposed in this work. For example, if taking
$N_\pi=N_\nu=5$ and $B=-0.2$, the numerical results of the eigenvalues are given as
\begin{eqnarray} E(0_1^+)=-28.0,~~E(2_1^+)=-27.2,~~E(4_1^+)=-26.0\,
.\end{eqnarray} These values exactly
coincide with the analytical formula given in (\ref{o6II}).
The eigenvectors are accordingly shown to be expanded in terms of
the SU(3) basis vectors with
\begin{eqnarray}\label{O6wave}
|\Psi\rangle=\sum_{\beta_\pi,\beta_\nu}C_{\beta_\pi,\beta_\nu}^L|N_\pi(\lambda_\pi,\mu_\pi)\chi_\pi
L_\pi;N_\nu(\lambda_\nu,\mu_\nu)\chi_\nu L_\nu;L\rangle\, ,
\end{eqnarray}
where $C_{\beta_\pi,\beta_\nu}^L$ represent the expansion
coefficients with the abbreviation
$\beta_{\rho}\equiv(\lambda_\rho\mu_\rho\chi_\rho L_\rho)$. In (\ref{O6wave}), the quantum number $M$ has been
ignored for convenience. Concretely, one can obtain
\begin{eqnarray}
\mid0_1^+\rangle=&-&0.4307\mid5(10,0)12;~5(10,0)12;~0\rangle\\
\nonumber &-&0.3340\mid5(10,0)10;~5(10,0)10;~0\rangle\\ \nonumber
&-&0.2556\mid5(6,2)22;~5(6,2)22;~0\rangle+\cdots,\\
\mid2_1^+\rangle=&&0.2076\mid5(10,0)12;~5(10,0)12;~2\rangle\\
\nonumber &-&0.2819\mid5(10,0)12;~5(10,0)10;~2\rangle\\ \nonumber
&-&0.1592\mid5(6,2)22;~5(6,2)22;~2\rangle+\cdots,\\
\mid4_1^+\rangle=&-&0.3220\mid5(10,0)12;~5(10,2)12;~4\rangle\\
\nonumber &-&0.2316\mid5(10,0)14;~5(10,0)10;~4\rangle\\ \nonumber
&-&0.1319\mid5(6,2)22;~5(6,2)22;~4\rangle+\cdots\, .
\end{eqnarray}
It is shown that each $\mathrm{O}_{\pi+\nu}(6)$ eigenvector could
be a complicated combination of the SU(3) basis vectors. As to octupole states, the
$\mathrm{O}_{\pi+\nu}(6)$ Hamiltonian (\ref{O6H}) may generate the
eigenvalues as same as those for quadrupole states, such as
$E(3_1^-)=E(0_1^+)$ and $E(2_1^-)=E(2_1^+)$, if keeping
$N_\pi=N_\nu=5$. This is actually the common features for any Hamiltonian
without involving $f$ boson operator (see Eq.~(\ref{quadrupoleoctupole})). Anyway, the example shows that the SU(3)
scheme works well in solving the IBM-2 Hamiltonian.

\subsection{The Model Hamiltonian}

To describe both positive-
and negative-parity states, the model Hamiltonian can be designed to
include three parts
\begin{eqnarray}\label{Hsdf}
\hat{H}=\hat{H}_{sd}+\hat{H}_f+\hat{V}_{sdf}\, .
\end{eqnarray}
The term $\hat{H}_{sd}$ describing the $sd$-boson core excitation is
responsible for both positive- and negative-parity states. The
latter two terms will appear only in the calculations for
negative-parity states. $\hat{H}_{sd}$ can be simply taken as the consistent-$Q$ form
\begin{eqnarray}\label{CQ}
&&\hat{H}_{sd}=\varepsilon_d(\hat{n}_{d_\pi}+\hat{n}_{d_\nu})+\kappa
\hat{Q}_\pi\cdot\hat{Q}_\nu\,
\end{eqnarray}
with
\begin{eqnarray}
&&\hat{n}_{d_\rho}=d_\rho^\dag\cdot\tilde{d}_\rho,~~~\rho=\pi,\nu\\
\label{qua}
&&\hat{Q}_\rho=(s_\rho^\dag\times\tilde{d}_\rho+d_\rho^\dag\times\tilde{s}_\rho)^{(2)}+\chi_\rho(d_\rho^\dag\times\tilde{d}_\rho)^{(2)}\,
\end{eqnarray}
and $\varepsilon_d$, $\kappa$ and $\chi_\rho$ representing the
interactional parameters. The $f$-boson Hamiltonian is given by
\begin{eqnarray}\label{Hf}
\hat{H}_f=\varepsilon_{f_\pi}\hat{n}_{f_\pi}+\varepsilon_{f_\nu}\hat{n}_{f_\nu}\,
\end{eqnarray}
with
\begin{eqnarray}
\hat{n}_{f_\rho}=-f_\rho^\dag\cdot\tilde{f}_\rho\,
\end{eqnarray}
and $\varepsilon_{f_\rho}$ denoting the single-$f_\rho$ boson energy. Here,
the other interactions between $f$ bosons have been ignored due to $n_f\leq1$. Such an approximation is equivalent to
the experimental constraint on the excitation energy
$E\ll2\varepsilon_{f_\rho}$~\cite{Scholten1978}. Clearly, a state involving two $f$ bosons must be
positive-parity one and is supposed to only appear at high energy in the present scheme.

For the interaction between $s~d$ and $f$ bosons, we emphasize here
the quadrupole-quadrupole and octupole-octupole
types~\cite{Barfield}. Specifically, it is given by
\begin{eqnarray}\label{Vsdf}\nonumber\hat{V}_{sdf}&=&t_{\pi\nu}\hat{Q}_{f_\pi}\cdot\hat{Q}_{\nu}+t_{\nu\pi}\hat{Q}_{f_\nu}\cdot\hat{Q}_{\pi}
+t_{\pi\pi}\hat{Q}_{f_\pi}\cdot\hat{Q}_{\pi}+t_{\nu\nu}\hat{Q}_{f_\nu}\cdot\hat{Q}_{\nu}\\
&~&+\omega_\pi\hat{\Omega}_\pi\cdot\hat{\Omega}_{\pi}+\omega_\nu\hat{\Omega}_\nu\cdot\hat{\Omega}_{\nu}+
\omega_{\pi\nu}\hat{\Omega}_\pi\cdot\hat{\Omega}_{\nu}\
\end{eqnarray}
with
\begin{eqnarray}
&\hat{Q}_{f_\rho}=-2\sqrt{7}(f_\rho^\dag\times\tilde{f}_\rho)^{(2)},\\
\label{oct}
&\hat{\Omega}_\rho=(s_\rho^\dag\times\tilde{f}_\rho+f_\rho^\dag\times\tilde{s}_\rho)^{(3)}
+\chi_{\rho3}(d_\rho^\dag\times\tilde{f}_\rho+f_\rho^\dag\times\tilde{d}_\rho)^{(3)}\,
,\end{eqnarray}
where $t_{\rho\rho^\prime}$, $\omega_\rho$ and
$\chi_{\rho3}$ ($\rho,\rho^\prime=\pi,\nu$) represent the adjustable parameters.
Some terms in (\ref{Vsdf}) may
contribute nothing in the calculations for $n_f=1$.
Nonetheless, there are still too many terms thus with many adjustable
parameters included in $\hat{V}_{sdf}$. If the type of $f$
boson can be specified in advance, the parameters in the
concrete calculations could be cut in half. For example, the
analysis given in \cite{Konljin1982} indicate that the low-lying
negative states in $^{152}$Sm may be dominated by the proton
octupole configuration, which means that the $f$ boson for this
nucleus can be fixed as $f_\pi$ when discussing the
low-energy structure in the present model. Accordingly, all the terms containing the $f_\nu$
boson operator can be ignored as an approximation.

\section{Application to $^{152}$Sm}

In this section, the critical point nucleus, $^{152}$Sm, will be
taken to test the IBM-2 description of both quadrupole and octupole states. Based on the above discussions, the Hamiltonian form for this nucleus is taken as
\begin{eqnarray}\label{Hsm}
\hat{H}=\hat{H}_{sd}+\varepsilon_{f_\pi}\hat{n}_{f_\pi}+t_{\pi\pi}\hat{Q}_{f_\pi}\cdot\hat{Q}_{\pi}+t_{\pi\nu}\hat{Q}_{f_\pi}\cdot\hat{Q}_{\nu}
\, ,
\end{eqnarray} where the $f$-boson bas been fixed as $f_\pi$. For simplicity, only the quadrupole-quadrupole
type interactions in $\hat{V}_{sdf}$ are considered here.
A similar Hamiltonian form but in the IBM-1 frame was ever used in the discussions \cite{Konljin1982,Scholten1978}.
To calculate the $B(E\lambda)$
transitions with $\lambda=1,2,3$, the transitional operators are
chosen as
\begin{eqnarray}
&&T^{(E1)}=e_1(d_\pi^\dag\times\tilde{f}_\pi+f_\pi^\dag\times\tilde{d}_\pi)^{(1)}\\
&&T^{(E2)}=e_2(\hat{Q}_\pi+\hat{Q}_\nu)\\
&&T^{(E3)}=e_3\hat{\Omega}_\pi\, ,\
\end{eqnarray}
in which $e_\lambda$ denote the effective charges.
In addition, one of the advantages of the IBM-2 against the IBM-1 consists in the description of the $M1$ properties~\cite{Iachellobook}, which can provide extra information to identify the nature of octupole states~\cite{Scheck2010}. In a recent work~\cite{Nomura2022}, the $sdf$-IBM-2 with $s,~d,~f$ bosons treated in equal footing has been applied to give a systematic study of the octupole correlations in several neutron-rich isotopes with some predictions of the $B(M1)$ transitional rates being provided in the calculations. To calculate the $B(M1)$ transitions, the $M1$ transitional operator will be chosen as same as that adopted in \cite{Nomura2022}. Specifically, it is defined by
\begin{eqnarray}
T^(M1)=\sqrt{\frac{3}{4\pi}}\sum_{\rho=\pi,\nu}g_\rho\hat{L}_\rho\,
\end{eqnarray}
with
\begin{equation}
\hat{L}_\rho=\sqrt{10}(d_\rho^\dag\times\tilde{d}_\rho)^{(1)}+\sqrt{28}(f_\rho^\dag+\tilde{f}_\rho)^{(1)}\, ,
\end{equation}
where $g_\rho$ represent the bosonic gyromagnetic (g) factors. In concrete calculations, the g factors are taken as $g_\pi=1$ $\mu_N$
and $g_\nu=0$ $\mu_N$~\cite{Iachellobook,Nomura2008}.

Counting from the valence shells $N_\mathrm{p}=50-82$
and $N_\mathrm{n}=82-126$, the boson number for $^{152}$Sm is
$N_\pi=6$ and $N_\nu=4$. Generally,
the model space of the IBM-2 is one order larger than that of the IBM-1 for a given nucleus. For example, the IBM-2 with $N_\pi+N_\nu=6+4=10$ will
produce 133 $0^+$ states and 414 $2^+$ states. In contrast, the IBM-1 with $N=10$ will generate 14 $0^+$ states and 22 $2^+$ states. If one
of the boson is assumed to be $f_\pi$, the number of the $2^-$ states will
reach $1344$ in the IBM-2+$f$ but only $54$ in the IBM-1+$f$. It means that there are more than $1.8\times10^6$ matrix elements
involved in the calculations for $2^-$ states.
In short, calculations in the IBM-2
could be much more time consuming than in the IBM-1 especially for the cases with
$f$ boson involved.

\begin{figure}
\begin{center}
\includegraphics[scale=0.41]{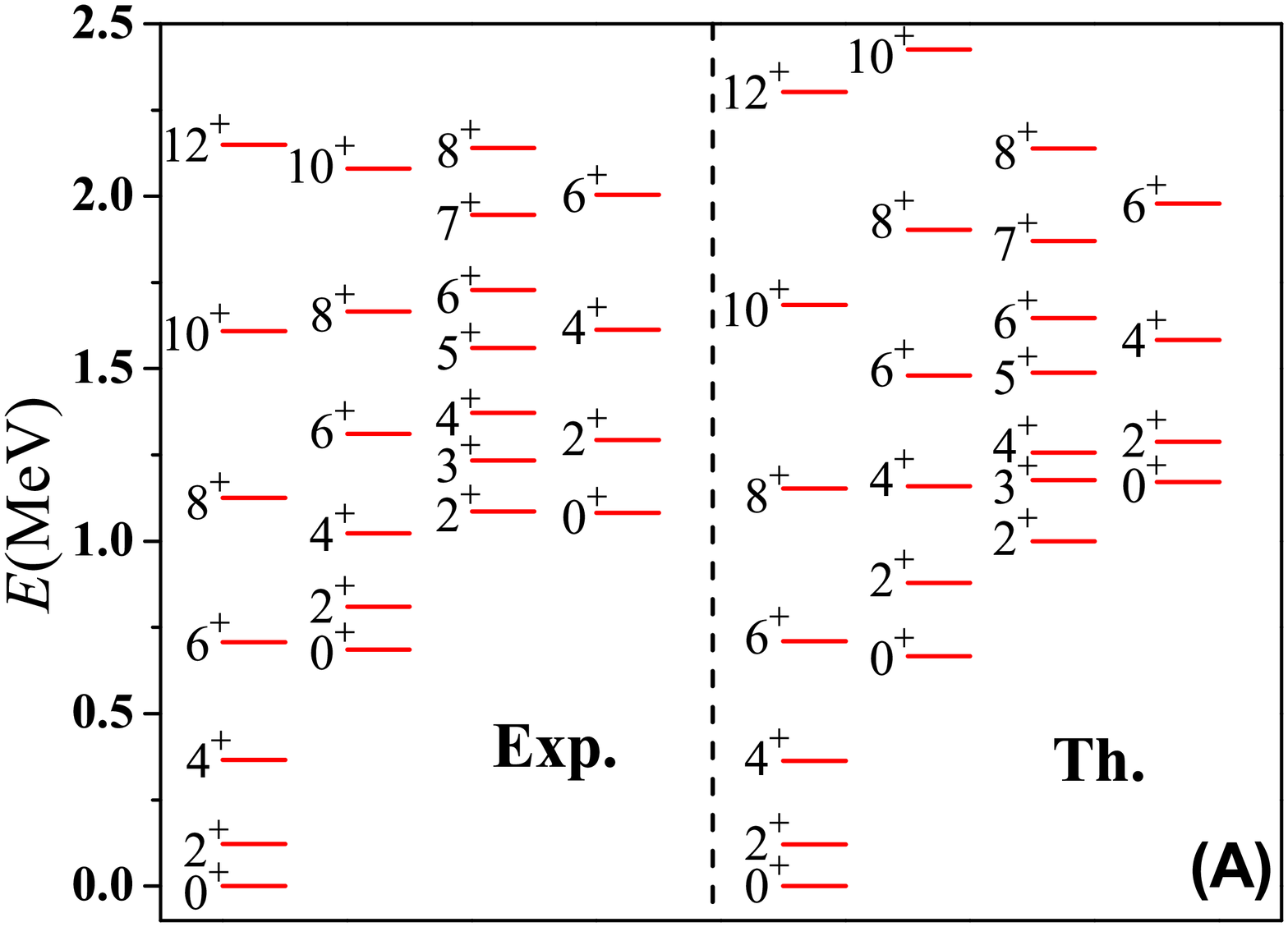}
\includegraphics[scale=0.41]{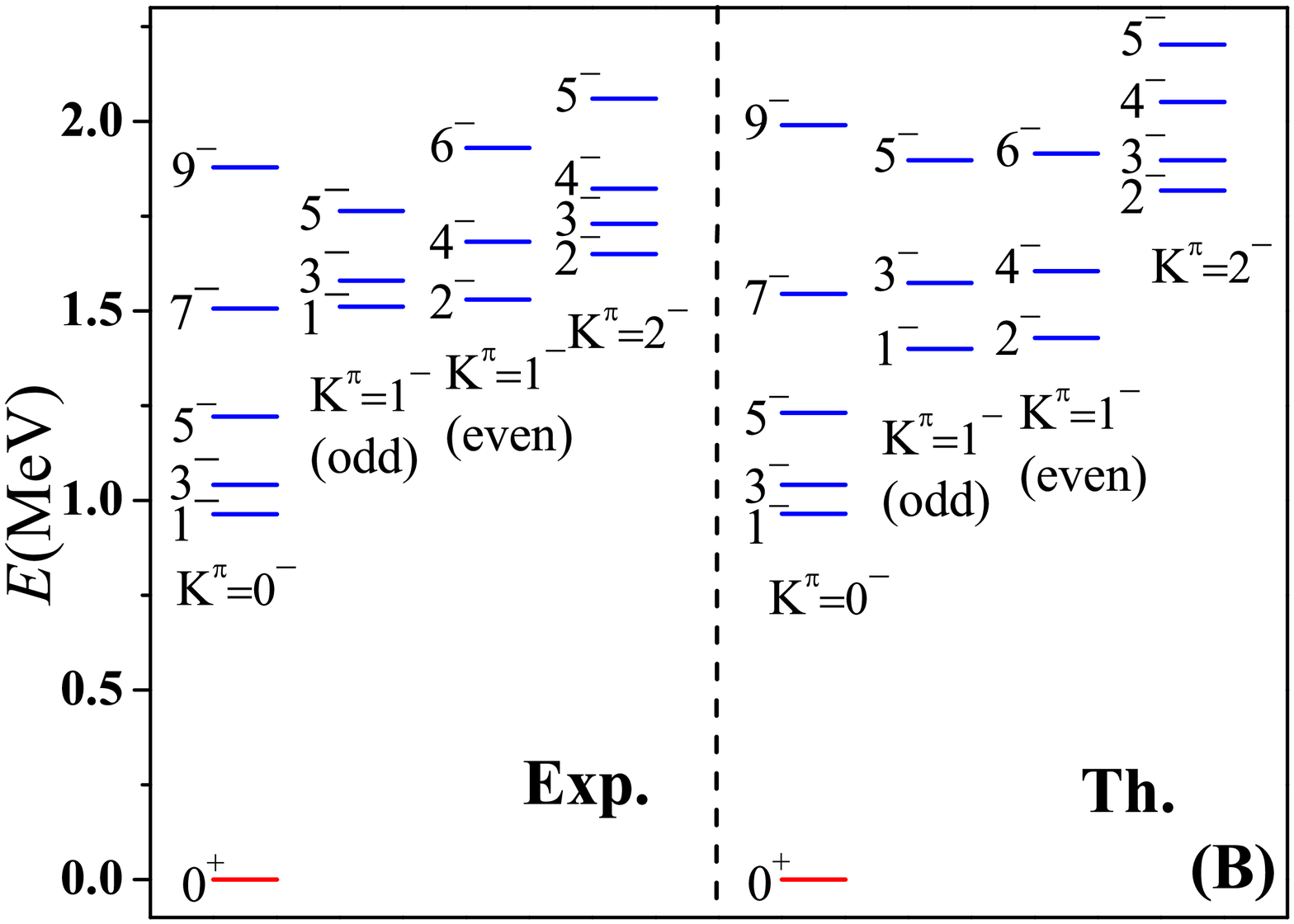}
\caption{Level patterns for both the positive-parity and negative-parity bands in
$^{152}$Sm with the data taken from \cite{Martin2013} are shown to compare with
the ones solved from (\ref{Hsm}). In the calculation, the parameters
are adopted as $\varepsilon_d=0.52$MeV, $\kappa=-0.074$MeV,
$\varepsilon_f=2.25$MeV, $t_{\pi\pi}=-0.043$MeV and
$t_{\pi\nu}=0.012$MeV, $\chi_\pi=-1.2$ and
$\chi_\nu=-1.0$.}\label{F1}
\end{center}
\end{figure}

\begin{figure}
\begin{center}
\includegraphics[scale=0.42]{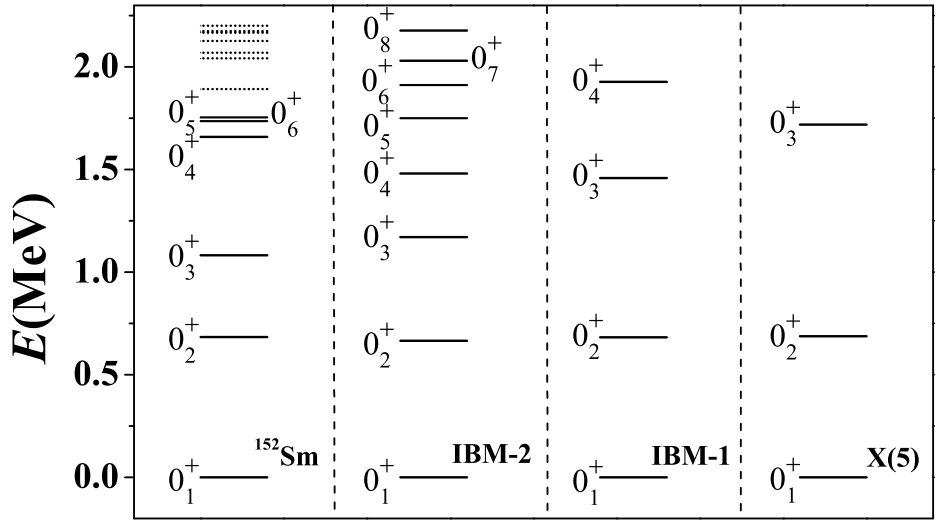}
\caption{The $0^+$ levels below 2.2MeV in $^{152}$Sm and those extracted from the IBM-2, IBM-1 and X(5) models.
The parameters in the IBM are taken as same as those adopted in Table~\ref{T2}, and the scale factor in X(5) has been
fixed from the data for $E(2_1^+)$. In addition, the dotted lines denote the possible $0^+$ levels
but with some uncertainties in the experiment~\cite{Martin2013}.}\label{F2}
\end{center}
\end{figure}

In Fig.~\ref{F1}, the level patterns bound below 2 MeV in $^{152}$Sm are
shown to compare with those solved from the
Hamiltonian (\ref{Hsm}). As seen in Fig.~\ref{F1}(A), the
level energies for the ground-state band,
$\gamma$-vibrational band, $\beta$-vibrational band and even second
$\beta$-vibrational band are all well reproduced by the
model calculations. For example, the band-head energy of the second $\beta$ band in experiment is
about $1.1$ MeV, which is in good agreement with
the present IBM-2 calculation with $E(0_3^+)\approx1.2$ MeV. In
contrast, the result fitted from the
consistent-$Q$ Hamiltonian in the IBM-1~\cite{Zhang2017} is given with
$E(0_3^+)\approx1.5$ MeV (see Fig.~\ref{F2}). Moreover,
the approximate degeneracy for $E(6_1^+)\approx E(0_2^+)$ in
the experiment, which was considered as a signature of the
spherical-deform shape shape phase transition~\cite{Bonatsos2008}, is
reproduced rather well by the IBM-2 calculation. As further seen in
FIG.~\ref{F1}(B), three negative-parity bands with $K^P=0^-,~1^-,~2^-$ in the
experiment are shown to compare with the theoretical calculations.
One may observe a reasonable agreement with the data for both band-head energies and level orders.
It means that the low-lying negative-parity states in $^{152}$Sm can indeed be
explored as octupole states. While, a discrepancy may come from the odd-even staggering presented in the calculations for the
$K^P=1^-$ band, which may be improved by involving other types of
interaction in the Hamiltonian.

\begin{table}
\caption{The available data (in W.u.) for the $B(E2)$, $B(E1)$, $B(E3)$ and $B(M1)$ transitions in $^{152}$Sm~\cite{Martin2013} with the unknown ones
denoted by "-" are shown to compare with the results solved from the IBM-2+$f$. In the calculations, the effective charges
(in $\sqrt{\mathrm{W.u.}}$) are adopted as $e_1=0.220$,
$e_2=2.142$ and $e_3=2.799$ with $\chi_{\pi3}=-1.0$. In addition, the $g$ factors (in $\mu_N$) have been set as $g_\pi=1$ and $g_\nu=0$.
Note that the symbol $J_n^\pi$ in each transition represents the initial or final state with the same level order as shown in
Fig.~\ref{F1}.}
\begin{center}
\begin{tabular}{ccc|ccc}
\hline\hline
&$B(E2)$&&&$B(E1)$ ($\times10^{-3}$)&\\
$J_i^\pi\rightarrow J_f^\pi$&Exp.&Th.&$J_i^\pi\rightarrow J_f^\pi$&Exp.&Th.\\ \hline
$2_1^+\rightarrow 0_1^+$&145.0(16)&145.0&$1_1^-\rightarrow 0_1^+$&5.8(5)&5.8\\
$4_1^+\rightarrow 2_1^+$&209.5(22)&213.5&$1_1^-\rightarrow 2_1^+$&10.6(9)&4.5\\
$6_1^+\rightarrow 4_1^+$&240(4)&236&$3_1^-\rightarrow 4_1^+$&8.2(16)&2.1\\
$8_1^+\rightarrow 6_1^+$&293(4)&242&$3_1^-\rightarrow 2_1^+$&8.1(15)&13.8\\
$10_1^+\rightarrow 8_1^+$&$314_{-26}^{+35}$&233&$1_2^-\rightarrow 2_1^+$&0.91(7)&2.32\\
$3_1^+\rightarrow 2_3^+$&$120_{-90}^{+60}$&142&$1_2^-\rightarrow 2_5^+$&0.172(25)&0.227\\
$0_2^+\rightarrow 2_1^+$&33.3(12)&40.8&$2_1^-\rightarrow 2_1^+$&0.26(5)&6.75\\
$0_3^+\rightarrow 2_1^+$&$0.8_{-23}^{+53}$&0.01&$2_1^-\rightarrow 2_3^+$&1.14(22)&0.21\\
$0_3^+\rightarrow 2_2^+$&$34_{-11}^{+23}$&33&$3_2^-\rightarrow 2_1^+$&0.255(22)&2.512\\
$2_2^+\rightarrow 2_1^+$&5.7(4)&12.3&$3_2^-\rightarrow 2_2^+$&0.29(3)&0.10\\
$2_3^+\rightarrow 2_1^+$&7.4(10)&2.4&$3_2^-\rightarrow 2_3^+$&0.40(4)&0.38\\
$4_3^+\rightarrow 2_3^+$&$62_{-24}^{+35}$&35&$3_2^-\rightarrow 3_1^+$&0.38(4)&0.01\\
$4_3^-\rightarrow 6_1^-$&$0.9_{-4}^{+6}$&0.6&$3_2^-\rightarrow 4_1^+$&1.26(11)&2.39\\
$1_2^-\rightarrow 3_1^-$&$3.6(7)$&5.8&$3_2^-\rightarrow 4_2^+$&0.163(15)&0.863\\
$2_1^-\rightarrow 1_1^-$&$1.4(9)$&7.0&$3_2^-\rightarrow 4_3^+$&1.30(12)&0.83\\
$2_1^-\rightarrow 3_1^-$&$25(5)$&13&$5_2^-\rightarrow 4_1^+$&$0.49_{-26}^{+49}$&2.31\\
$3_2^-\rightarrow 1_1^-$&$8.1(9)$&9.2&$5_2^-\rightarrow 6_1^+$&$1.4_{-8}^{+14}$&1.1\\
\hline\hline
&$B(E3)$&&&$B(M1)$&\\
$J_i^\pi\rightarrow J_f^\pi$&Exp.&Th.&$J_i^\pi\rightarrow J_f^\pi$&Exp.&Th.\\ \hline
$3_1^-\rightarrow 0_1^+$&13.9(15)&13.9&$2_2^+\rightarrow 2_1^+$&$0.000015(7)$&0.016591\\
$3_2^-\rightarrow 0_1^+$&7.3(21)&5.30&$2_3^+\rightarrow 2_1^+$&0.00015(3)&0.009868\\
$3_3^-\rightarrow 0_1^+$&-&0.12&$2_3^+\rightarrow 2_2^+$&0.00134(18)&0.008284\\
$1_1^-\rightarrow 2_1^+$&-&25.32&$3_1^+\rightarrow 2_1^+$&$0.00021_{-5}^{+6}$&0.001918\\
$1_1^-\rightarrow 4_1^+$&-&14.41&$3_1^+\rightarrow 2_3^+$&$0.0051_{-38}^{+25}$&0.002275\\
$3_1^-\rightarrow 2_1^+$&-&15.02&$3_1^+\rightarrow 4_1^+$&$0.00024_{-5}^{+6}$&0.002283\\
$3_1^-\rightarrow 4_1^+$&-&9.20&$4_2^+\rightarrow 4_1^+$&$0.00090(25)$&0.031915\\
$5_1^-\rightarrow 2_1^+$&-&22.02&$4_3^+\rightarrow 4_1^+$&$0.0014_{-6}^{+8}$&0.069329\\
$5_1^-\rightarrow 4_1^+$&-&12.51&$2_1^-\rightarrow 1_1^-$&0.0015(6)&0.000149\\
$5_1^-\rightarrow 6_1^+$&-&5.70&$2_1^-\rightarrow 3_1^-$&$0.00036_{-9}^{+10}$&$0.000021$\\
\hline\hline
\end{tabular}\label{T1}
\end{center}
\end{table}

To further check the model, the available data and the theoretical results for the
$B(E1)$, $B(E2)$, $B(E3)$ and $B(M1)$ transitions are listed in
Table~\ref{T1}. As observed from Table~\ref{T1}, the $B(E2)$ transition values
in $^{152}$Sm are in good agreement with the model predictions especially for the big differences between the interband and intraband
transitions, which in turn confirms the band assignments shown in Fig.~\ref{F1}. Similarly, the data for the $B(E1)$ transitions can be also
reasonably explained by the model calculations. But, some $B(E1)$ values in the experiment, such as $B(E1;2_1^-\rightarrow2_3^+)$,
are larger than the calculated results, by which a closer relation between the $K^P=1^-$ band and $\gamma$ band for low-spin states is thus suggested. As for $B(E3)$
transition, although two data are available, the strong $E3$ transition with $B(E3;3_1^-\rightarrow 0_1^+)$ support that the $K^P=0^-$ band may originate from an octupole vibration based upon the ground state $0_1^+$~\cite{Konljin1982}.
Overall, the consistency between the experimental data and the theoretical calculations for the $B(E\lambda)$ transitions
confirm that the low-lying negative-parity structure in $^{152}$Sm should be dominated by the octupole configurations. It is worth mentioning that the octupole states in this nucleus were ever discussed within
the framework of IBM-1+$f$~\cite{Scholten1978,Konljin1982}. But only two negative-parity bands were involved in the previous discussions. The results indicate that the IBM-1+$f$ (one-fluid model) and the IBM-2+$f$ (two-fluid model)
may give the qualitatively similar descriptions of the two lowest negative-parity bands. Nevertheless, one advantage of two-fluid model against one fluid model is that the $M1$ properties can be directly calculated using one body transitional operator~\cite{Iachellobook}. One may observe from Table~\ref{T1} that the experimental data for the $B(M1)$ transition can be reasonably explained from the IBM-2 calculations too, which confirms again that the IBM-2 can provide a better frame to describe more states of both positive-parity and negative-parity in $^{152}$Sm.

\begin{table}
\caption{Typical energy ratios and $B(E2)$ ratios for quadrupole states in $^{152}$Sm are shown to compare with those extracted from the IBM-2 and IBM-1 calculations as well as the results predicted from the X(5) critical point symmetry~\cite{Iachello2001}. In the calculations, the IBM-2 parameters are taken as the same ones adopted in Fig.~\ref{F1}, while the IBM-1 parameters are taken from \cite{Zhang2017}.}
\begin{center}
\begin{tabular}{ccccc|ccccc}
\hline\hline
Ratios&Exp.&IBM-2&IBM-1&X(5)&Ratios&Exp.&IBM-2&IBM-1&X(5)\\ \hline
$\frac{E(4_1^+)}{E(2_1^+)}$&3.01&2.99&3.02&2.90&$\frac{B(E2;4_1^+\rightarrow 2_1^+)}{B(E2;2_1^+\rightarrow 0_1^+)}$&1.44&1.47&1.47&1.58\\
$\frac{E(6_1^+)}{E(2_1^+)}$&5.80&5.86&5.93&5.43&$\frac{B(E2;6_1^+\rightarrow 4_1^+)}{B(E2;2_1^+\rightarrow 0_1^+)}$&1.66&1.56&1.62&1.98\\
$\frac{E(0_2^+)}{E(2_1^+)}$&5.62&5.49&5.60&5.65&$\frac{B(E2;0_1^+\rightarrow 2_1^+)}{B(E2;2_1^+\rightarrow 0_1^+)}$&0.23&0.28&0.28&0.83\\
$\frac{E(2_2^+)}{E(2_1^+)}$&6.66&7.25&7.71&7.45&$\frac{B(E2;2_2^+\rightarrow 2_1^+)}{B(E2;2_1^+\rightarrow 0_1^+)}$&0.04&0.08&0.07&0.09\\
$\frac{E(0_3^+)}{E(2_1^+)}$&8.89&9.65&11.98&14.12&$\frac{B(E2;0_3^+\rightarrow 2_2^+)}{B(E2;2_1^+\rightarrow 0_1^+)}$&0.23&0.23&0.56&0.85\\
\hline\hline
\end{tabular}\label{T2}
\end{center}
\end{table}

As is known, quadrupole states in $^{152}$Sm have been extensively discussed by different models.
In Table~\ref{T2}, we make a close comparison between three models by calculating some typical ratios for quadrupole states. One can find
from Table~\ref{T2} that the IBM-2, IBM-1 and X(5) models can all give good reproductions of the experimental values in $^{152}$Sm. Among them, the X(5) results are even obtained from the parameter-free predictions, which thus provide a benchmark for
the critical point structure in the U(5)-SU(3) transition~\cite{Iachello2001}. However, the level energy $E(0_3^+)$
is obviously overestimated by both the IBM-1 and X(5) calculations. In contrast, the IBM-2 can accommodate more excited $0^+$
states due to its larger model space than the IBM-1. To further emphasize this point, the $0^+$
levels up to 2.2 MeV in $^{152}$Sm~\cite{Martin2013} are shown in Fig.~\ref{F2}
to make an additional comparison with the theoretical calculations. As seen from Fig.~\ref{F2}, there are more than 6 levels with $J^\pi=0^+$ bound below 2.2 MeV in the experiment. Apart from a quantitatively good agreement for the lowest ones, the distribution of these experimental $0^+$ levels can be roughly understood from the IBM-2 calculations, such as a rapid enhancement in the level density above 1.5 MeV. In contrast, there are totally 3-4 $0^+$ levels bound below 2.2 Mev in the X(5) or in the IBM with only
the lowest two ones being consistent with the experimental results. According to the analysis in \cite{Meyer2006}, the emergence of many low-energy excited $0^+$ states can be a signature of the spherical-deformed shape phase transitions in experiments.
Another way to generate more low-lying $0^+$ states in the IBM-1 frame is to involve more $f$ bosons. The resulting $sdf$-IBM-1 model has been employed to describe the rare-earth nuclei~\cite{Nomura2014,Nomura2015}. Many low-energy $0^+$ states in this model are shown to be attributed to the coupling of two-octupole phonons~\cite{Nomura2015}. Therefore, the true nature of the $0^+$ states in a transitional system could be more complicated than expected. On the other side, the distribution of $0^+$ states can be used to
analyze statistical properties of a nuclear system. Such kinds
of analysis were extensively performed in the framework of IBM-1~\cite{Alhassid1990,Alhassid1992,Alhassid1993,Whelan1993,Karampagia2015,Zhang2021},
which is used to simulate a large-$N$ nuclear system.
Undoubtedly, the IBM-2 can provide a better simulation of nuclear
systems.

\begin{figure}
\begin{center}
\includegraphics[scale=0.18]{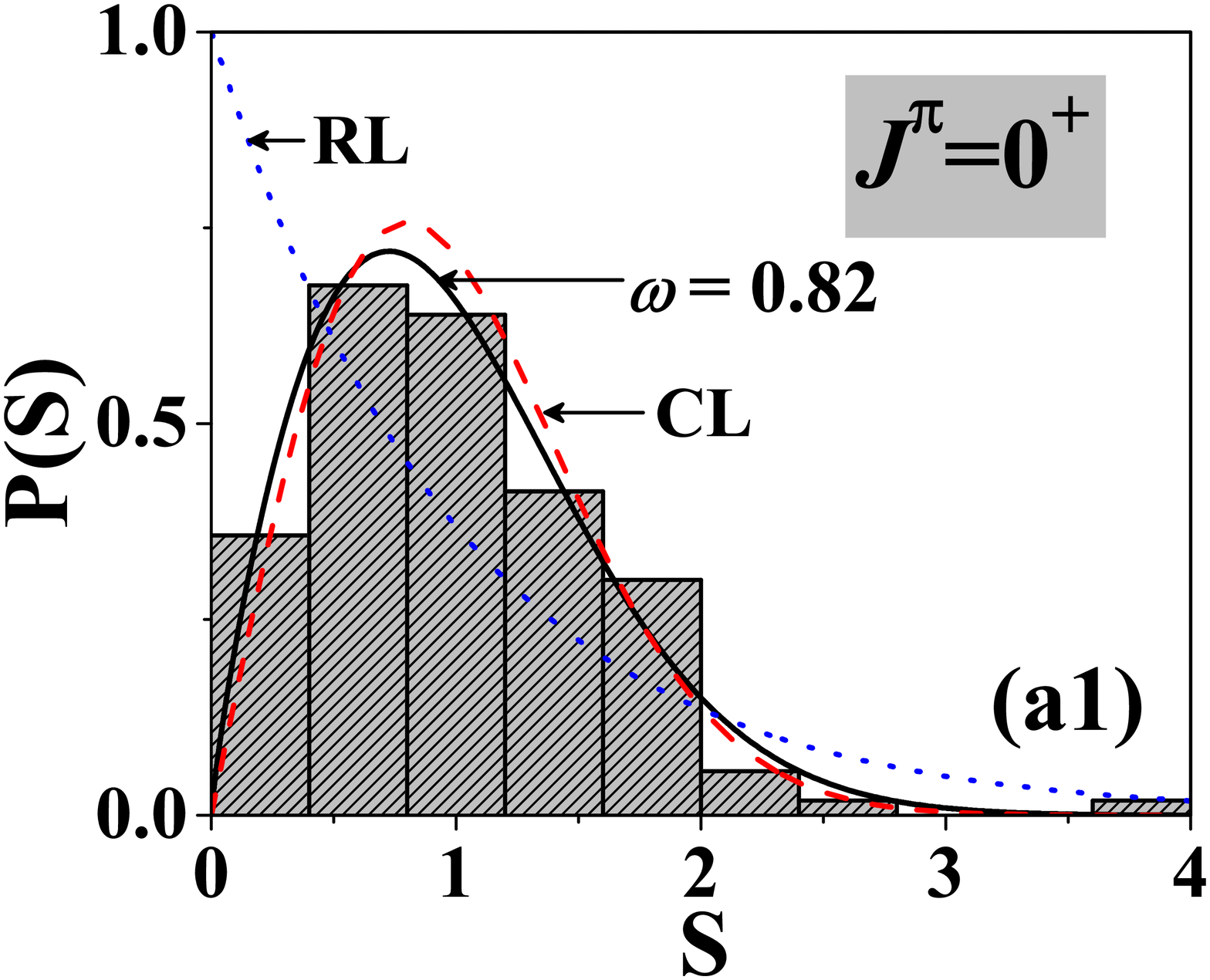}
\includegraphics[scale=0.18]{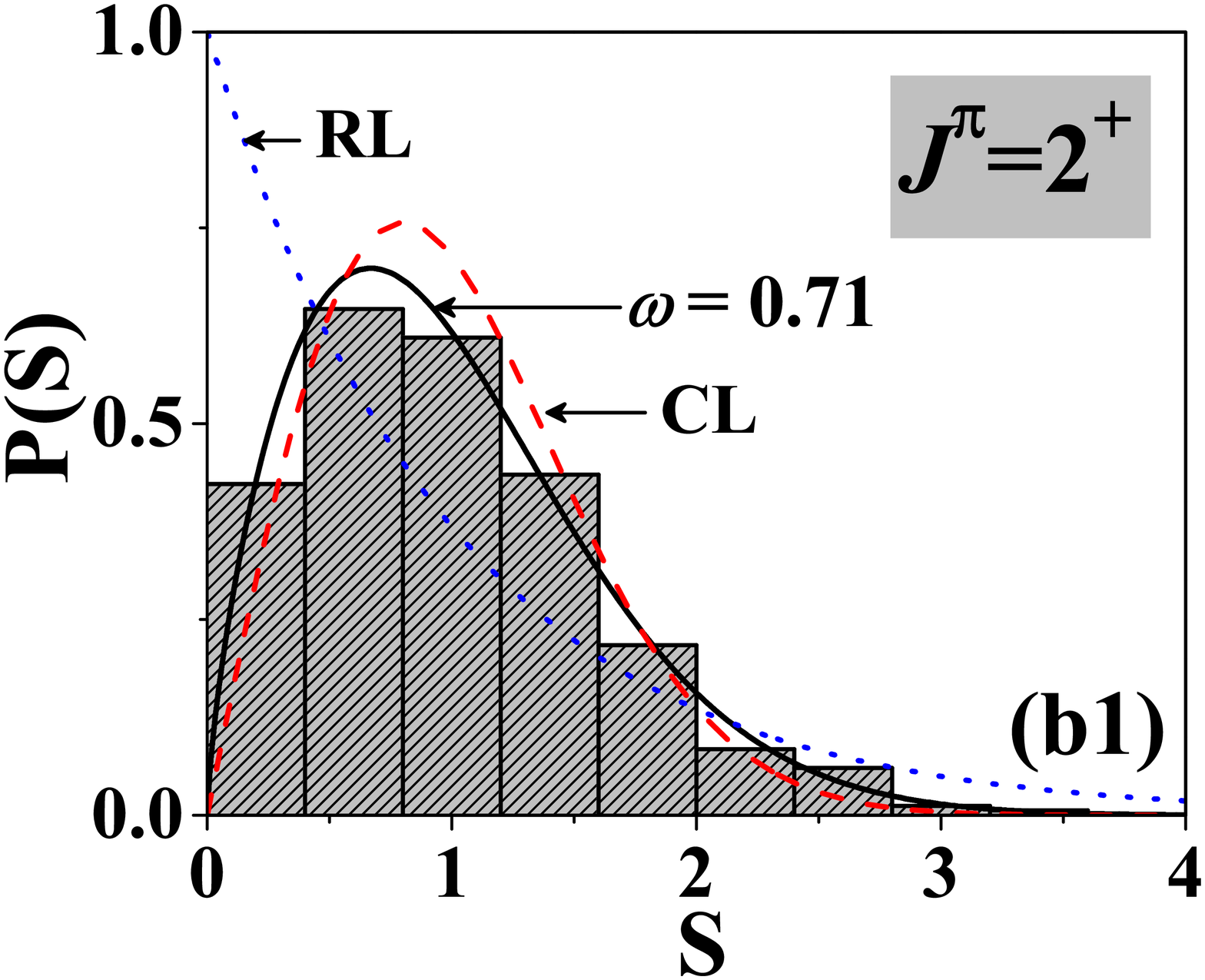}
\includegraphics[scale=0.18]{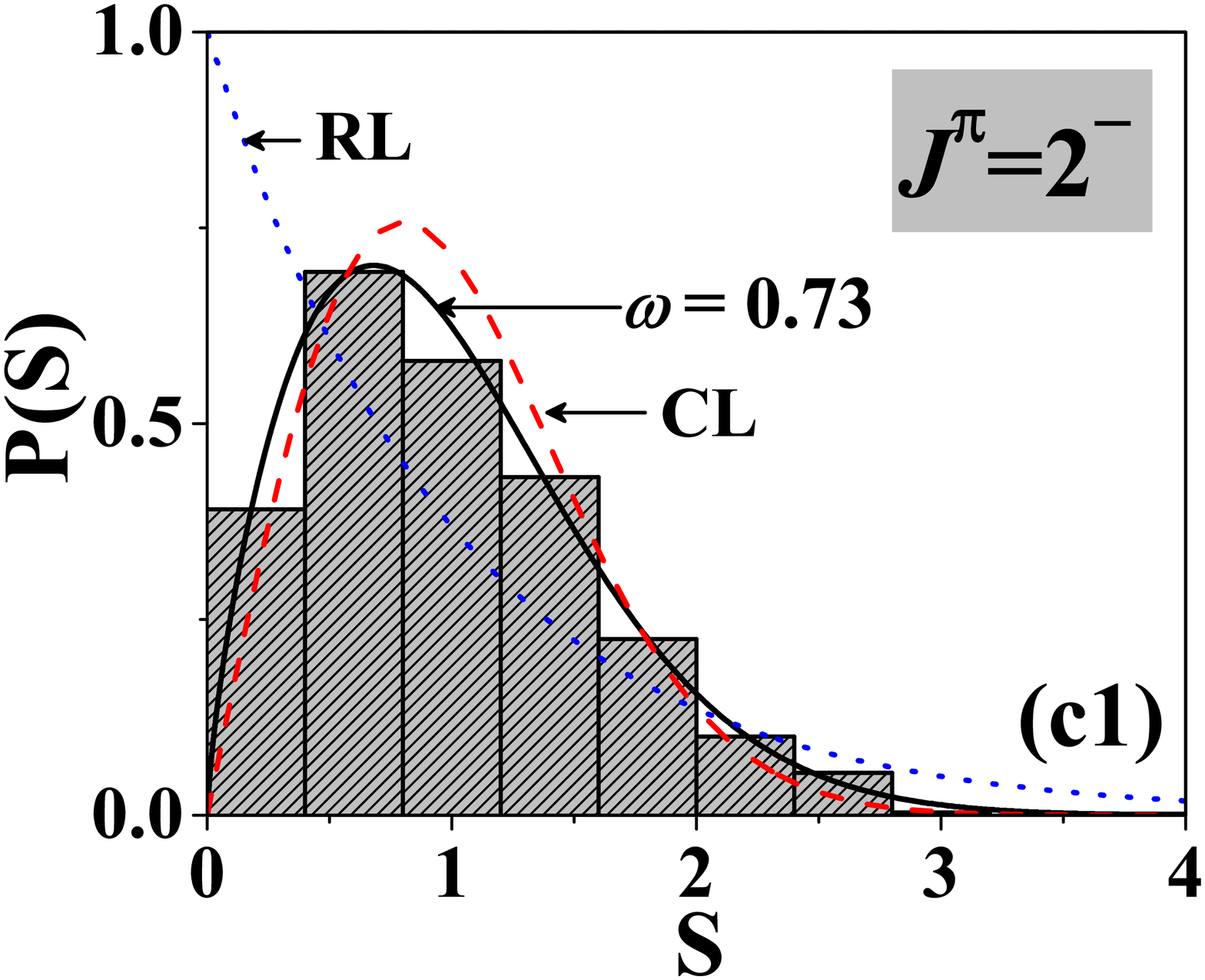}
\includegraphics[scale=0.18]{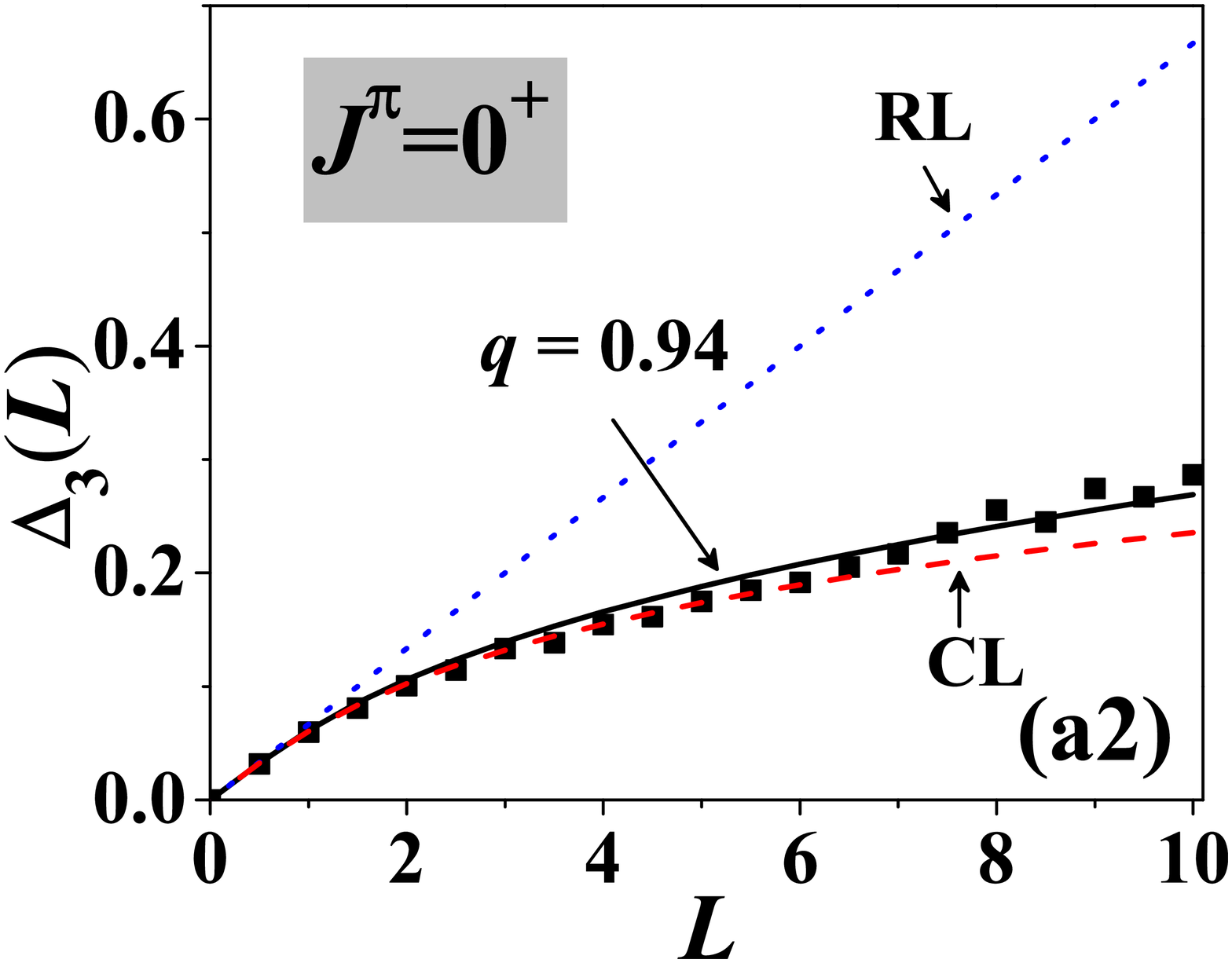}
\includegraphics[scale=0.18]{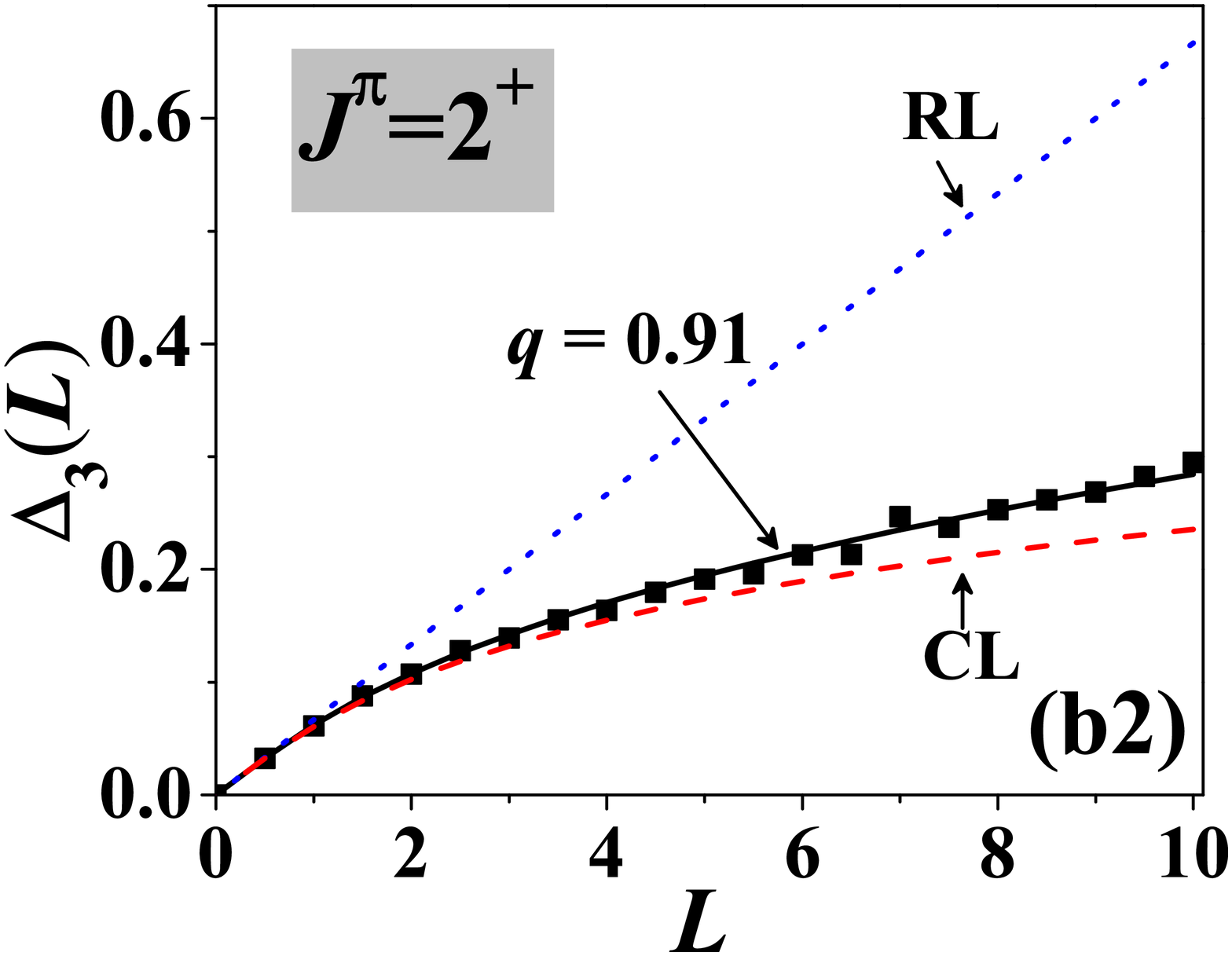}
\includegraphics[scale=0.18]{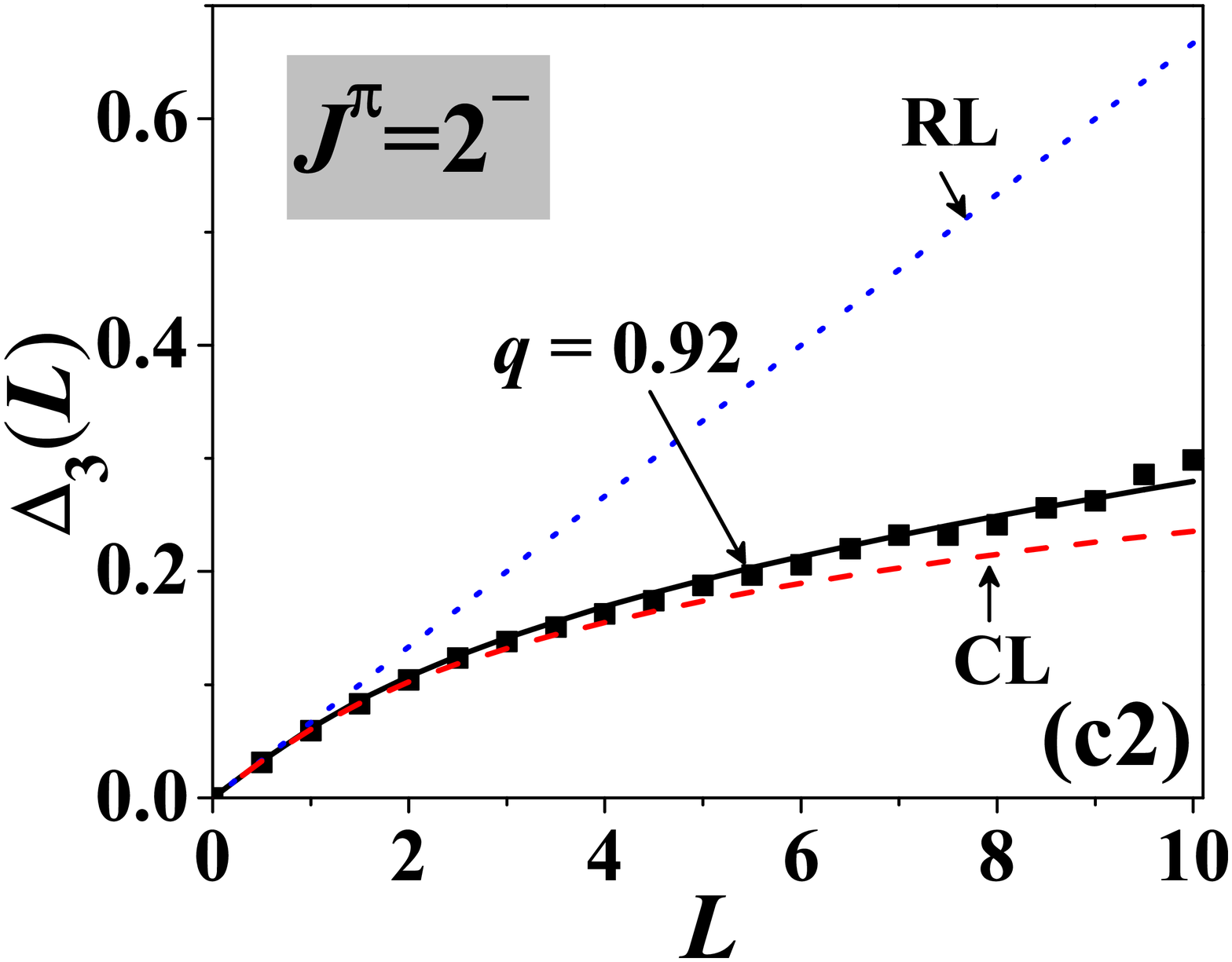}
\caption{The $P(S)$ and $\Delta_3(L)$ statistics on the
$0^+,~2^+,~2^-$ spectra that are solved from the Hamiltonian (\ref{Hsm}) for
$^{152}$Sm.  RL and CL represent the statistical results for the regular and chaotic limits,
respectively.}\label{F3}
\end{center}
\end{figure}

In contrast to the IBM-1, where an unrealistically large number
of boson has to be considered for the statistical
calculations~\cite{Alhassid1990,Alhassid1992,Alhassid1993,Whelan1993,Karampagia2015,Zhang2021}, the
IBM-2 with a realistic boson number may contain enough number of
states to do the same statistical analysis. To test spectral fluctuations
in a transitional system., we carry out the
spectral statistics on the low-spin states that are solved from the
IBM-2 Hamiltonian for $^{152}$Sm.  In the statistics, two quantum
measures, the nearest neighbor level spacing distribution $P(S)$ and
the $\Delta_3(L)$ statistics of Dyson and Mehta, are
adopted~\cite{Whelan1993}. Two parameters, $\omega$ and $q$,
are further used to fit the measures in order to estimate the possible deviation of the
system from the regular limit ($\omega=0$, $q=0$) or chaotic limit
($\omega=1$, $q=1$). For more details on how to calculate the two quantum
measures in the IBM, one can refer to~\cite{Karampagia2015,Zhang2021}.
In Fig.~\ref{F3}, the $P(S)$ and $\Delta_3(L)$
results are shown for the spectra with different $J^\pi$.
One can find from Fig.~\ref{F3} that the $P(S)$ statistics on the $0^+,~2^+$
and $2^-$ spectra give $\omega=0.82$, $\omega=0.71$ and
$\omega=0.73$, respectively. It means that the transitional system
described by (\ref{Hsm}) is chaotic thus with large spectral
fluctuations. This point actually agrees with the statistical analysis of the transitional Gd nuclei given in \cite{Meyer2006}.
The spectral chaos in the transitional system is even highlighted by the $\Delta_3(L)$ statistics, to
which the fitted $q$ values for the $0^+,~2^+$ and $2^-$ spectra are all given with $q>0.9$. As shown in Fig.~\ref{F1}, the
low-lying states in both the experiment and theory can be assigned into rotational bands,
which means that the low-lying dynamics in the transitional system should be less chaotic. It is thus deduced that the spectral chaos
for a given $J^\pi$ may mainly come from the high-lying ones, which is actually consistent with the statistical analysis of the IBM-1~\cite{Zhang2021}. In addition, the X(5) model can also give a good description of the low-lying properties in
the transitional system as discussed above. Nevertheless, the previous statistical analysis of the X(5) model~\cite{Hou2009} showed that the critical point symmetry may correspond to a rather regular situation with very small spectral fluctuations. In other words, the X(5) model exhibits a nearly inverse picture of the IBM-2 (as indicated in Fig.~\ref{F3}) from the point of view of spectral statistics. Therefore, the critical models with similar low-energy structures may have completely different dynamical features at high energy.

\section{Summary}

In summary, the IBM-2 with one $f$-boson coupled has been applied
to describe quadrupole and octupole states in a
shape-phase transitional system. To solve the model, a new scheme of building the
Hamiltonian matrix in terms of the weak-coupling SU(3)
basis is described in details. A further application to $^{152}$Sm
indicate that the low-lying properties of both the positive-parity (quadrupole) and
negative-parity (octupole) bands in the experiment
can be reasonably explained in the IBM-2+$f$ model through a simple
Hamiltonian form. Particularly, the IBM-2 due to
its richer model configurations can accommodate more $0^+$ states in experiments than
the IBM-1 and X(5) models. It is thus concluded that the IBM-2 with a robust microscopic basis~\cite{IachelloTalmi,Otsuka1978,Otsuka1981,Otsuka1978II}
can provide a better frame in a unified description of the quadrupole-octupole collectivity in a
transitional system, which may inspire us to further apply the model to explore the
other aspects in transitional nuclear structures, such as those
exhibited in nucleon-transfer or $\beta$-decay reactions.
In addition, a statistical analysis of the IBM-2
is also provided here. The results indicate that the vibrational spectra in the transitional system at low spins could be rather chaotic.
However, whether or not the chaotic features can appear in the other dynamical cases requires a systematically
statistical analysis of the IBM-2. Related work is in progress.

\section*{acknowledgments}
We wish to thank Z. Liu for stimulating this work and F. Pan for
many useful discussions on the octupole model. Support from the National
Natural Science Foundation of China (11875158) is acknowledged.


\section*{References}

\begin{thebibliography}{90}

\bibitem{Iachellobook}F. Iachello and A. Arima, {\it The Interacting Boson
Model} (England: Cambridge University, 1987)

\bibitem{IachelloTalmi}F. Iachello and I. Talmi, Rev. Mod. Phys. {\bf 59}, 339 (1987).

\bibitem{Otsuka1981}T. Otsuka, Phys. Rev. Lett. {\bf 46}, 710 (1981).

\bibitem{Otsuka1978}T. Otsuka, A. Arima, F. Iachello, and I. Talmi, Phys. Lett. B {\bf 76}, 139 (1978).

\bibitem{Otsuka1978II}T. Otsuka, A. Arima, and F. Iachello, Nucl. Phys. A {\bf 309}, 1 (1978).

\bibitem{Nomura2008}K. Nomura, N. Shimizu, and T. Otsuka, Phys. Rev. Lett. {\bf 101}, 142501 (2008).

\bibitem{Nomura2011}K. Nomura, T. Otsuka, R. Rodr{\'i}guez-Guzm{\'a}n, L. M. Robledo, and P. Sarriguren, Phys. Rev. C {\bf 83}, 014309 (2011).

\bibitem{Nomura2011II}K. Nomura, T. Otsuka, R. Rodr{\'i}guez-Guzm{\'a}n, L. M. Robledo, P. Sarriguren, P. H. Regan, P. D. Stevenson, and Zs. Podoly{\'a}k, Phys. Rev. C {\bf 83}, 054303 (2011).

\bibitem{Nomura2011III}K. Nomura, T. Otsuka, R. Rodr{\'i}guez-Guzm{\'a}n, L. M. Robledo, and P. Sarriguren, Phys.
Rev. C {\bf 84}, 054316 (2011).

\bibitem{Nomura2011IV}K. Nomura, T. Otsuka, N. Shimizu, and L. Guo, Phys. Rev. C {\bf 83}, 041302(R) (2011).

\bibitem{Nomura2017}K. Nomura, R. Rodr{\'i}guez-Guzm{\'a}n, and L. M.
Robledo, Phys. Rev. C {\bf 95}, 064310 (2017).

\bibitem{Nomura2019}K. Nomura and Y. Zhang. Phys. Rev. C {\bf 99}, 024324 (2019).


\bibitem{Barea2012}J. Barea, J. Kotila, and F. Iachello, Phys. Rev. Lett. {\bf 109}, 042501 (2012).

\bibitem{Barea2013}J. Barea, J. Kotila, and F. Iachello, Phys. Rev. C {\bf 87}, 014315 (2013).

\bibitem{Martin2013}M. J. Martin, Nucl. Data Sheets {\bf 114}, 1497 (2013).


\bibitem{Iachello1998}F. Iachello, N. V. Zamfir, and R. F. Casten, Phys. Rev. Lett. {\bf 81}, 1191 (1998).

\bibitem{Iachello2001}F. Iachello, Phys. Rev. Lett. {\bf 87}, 052502 (2001).

\bibitem{Casten2001}R. F. Casten and N. V. Zamfir, Phys. Rev. Lett. {\bf 87}, 052503 (2001).

\bibitem{Jolie1999}J. Jolie, P. Cejnar, and J. Dobe{\v s}, Phys. Rev. C {\bf 60}, 061303(R) (1999).

\bibitem{Li2009}Z. P. Li, T. Nik{\v s}i{\'c}, D. Vretenar, J. Meng, G. A. Lalazissis, and P.
Ring, Phys. Rev. C {\bf 79}, 054301 (2009).

\bibitem{Gupta2017}J. B. Gupta and J. H. Hamilton, Phys. Rev. C {\bf 96}, 034321 (2017).

\bibitem{Heyde2011}K. Heyde and J. L. Wood, Rev. Mod. Phys. {\bf 83}, 1467 (2011).

\bibitem{CJC2010}P. Cejnar, J. Jolie, and R. F. Casten, Rev. Mod. Phys. {\bf 82}, 2155 (2010).

\bibitem{Caprio2004}M. A. Caprio and F. Iachello, Phys. Rev. Lett. {\bf 93}, 242502 (2004).


\bibitem{Kotila2012}J. Kotila, K. Nomura, L. Guo, N. Shimizu, and T. Otsuka, Phys. Rev. C {\bf 85}, 054309 (2012).

\bibitem{Konljin1982}J. Konljn, J. B. R. Berkhout, W. H. A. Hesselink, J. J. Van Ruijven, P. Van Nes, H. Verheul,
F. W. N. De Boer, C. A. Fields, E. Sugarbaker, P. M. Walker, and R. Bijker, Nucl. Phys. A {\bf 373}, 397 (1982).

\bibitem{Yoshinaga1993}N. Yoshinaga, T. Mizusaki, and T. Otsuka, Nucl. Phys. A {\bf 559}, 193 (1993).

\bibitem{Smirnova2000}N. A. Smirnova, N. Pietralla, T. Mizusaki, and P. Van Isacker, Nucl. Phys. A {\bf 678}, 235 (2000).

\bibitem{Pietralla2003}N. Pietralla, C. Fransen, A. Gade, N. A. Smirnova, P. von Brentano, V. Werner, and S. W. Yates,
Phys. Rev. C {\bf 68}, 031305(R) (2003).

\bibitem{Nomura2022}K. Nomura, Phys. Rev. C {\bf 105}, 054318 (2022).


\bibitem{Vallejos2021}O. Vallejos and J. Barea, Phys. Rev. C {\bf 104}, 014308 (2021).

\bibitem{Scholten1978} O. Scholten, F. Iachello, and A. Arima, Ann. Phys. {\bf 115}, 325 (1978).

\bibitem{Barfield1986} A. F. Barfield, J. L. Wood, and B R. Barrett, Phys. Rev. C {\bf 34}, 2001
(1986).
\bibitem{Barfield1988} A. F. Barfield, B. R. Barrett, J. L. Wood, and O. Scholten, Ann. Phys. {\bf 182}, 344 (1988).

\bibitem{Engel1987} J. Engel and F. Iachello, Nucl. Phys. A {\bf 472}, 61 (1987).

\bibitem{Otsuka1988} T. Otsuka and M. Sugita, Phys. Lett. B {\bf 209}, 140 (1988).

\bibitem{Zamfir2001} N. V. Zamfir and D. Kusnezov, Phys. Rev. C {\bf 63}, 054306 (2001).

\bibitem{Nomura2013} K. Nomura, D. Vretenar, and B. N. Lu, Phys. Rev. C {\bf 88}, 021303(R) (2013)


\bibitem{Nomura2014} K. Nomura, D. Vretenar, T. Nik{\v s}i{\'c}, and B. N. Lu, Phys. Rev. C {\bf 89}, 024312 (2014)

\bibitem{Nomura2015}K. Nomura, R. Rodr{\'i}guez-Guzm{\'a}n, and L. M. Robledo, Phys. Rev. C {\bf 92}, 014312 (2015).



\bibitem{NPBOS}T. Otsuka and N. Yoshida, User's manual of the program NPBOS.
(1985).

\bibitem{DraayerAkiyama1973} J. P. Draayer and Y. Akiyama, J. Math.
Phys. {\bf 14}, 1904 (1973).


\bibitem{Barfield} A. F. Barfield, Ph. D. thesis, University of Arizona, (1986).


\bibitem{Rosensteel1990}G. Rosensteel, Phys. Rev. C {\bf 41}, 730, (1990).

\bibitem{Scheck2010}M. Scheck, P. A. Butler, C. Fransen, V. Werner, and S. W. Yates, Phys. Rev. C {\bf 81}, 064305 (2010).


\bibitem{Zhang2017}Y. Zhang and F. Iachello, Phys. Rev. C {\bf 95}, 034306 (2017).

\bibitem{Bonatsos2008}D. Bonatsos, E. A. McCutchan, R. F. Casten, and R. J. Casperson, Phys. Rev. Lett. {\bf 100}, 142501 (2008).

\bibitem{Meyer2006}D. A. Meyer {\it et.~al.}, Phys. Rev. C {\bf 74}, 044309 (2006).

\bibitem{Alhassid1990}Y. Alhassid, A. Novoselsky, and N. Whelan, Phys. Rev. Lett. {\bf 65}, 2971  (1990).

\bibitem{Alhassid1992}Y. Alhassid and A. Novoselsky, Phys. Rev. C {\bf 45}, 1677 (1992).

\bibitem{Alhassid1993}Y. Alhassid and N. Whelan, Phys. Rev. Lett. {\bf 70}, 572 (1993).

\bibitem{Whelan1993}N. Whelan and Y. Alhassid, Nucl. Phys. A {\bf556}, 42 (1993).

\bibitem{Karampagia2015}S. Karampagia, D. Bonatsos, and R. F. Casten, Phys. Rev. C {\bf 91}, 054325 (2015).

\bibitem{Zhang2021}W. T. Dong, Y. Zhang, B. C. He, F. Pan, Y. A. Luo, J. P. Draayer,
and S. Karampagia, J. Phys. G {\bf 48}, 045103 (2021).



\bibitem{Hou2009}Z. F. Hou, Y. Zhang, and Y. X. Liu, Phys. Rev. C {\bf 80}, 054308 (2009).


\end{thebibliography}
\end{document}